\documentclass[journal]{IAENGtran}
\usepackage{bm, amsmath}
\usepackage{color}
\usepackage{listings}    

\definecolor{dkgreen}{rgb}{0,0.6,0}
\definecolor{BlueDeFrance}{rgb}{0.19,0.55,0.91}
\definecolor{MyDarkBlue}{rgb}{0.0,0,0.7}
\definecolor{mygray}{rgb}{0.95,0.95,0.95}
\definecolor{brown}{rgb}{0.59,0.29,0}
\ifCLASSINFOpdf
   \usepackage[pdftex]{graphicx}
   \DeclareGraphicsExtensions{.pdf,.jpeg,.png}
\else
   \usepackage[dvips]{graphicx}
   \DeclareGraphicsExtensions{.eps}
\fi

\begin{document}
%
\title{Python Classes for Numerical Solution of PDE's}
%
%
%

\author{Asif~Mushtaq,~\IAENGmembership{Member,~IAENG,}
	Trond~Kvamsdal,
	K{\aa}re Olaussen,~\IAENGmembership{Member,~IAENG,} %
\thanks{Manuscript received January 22, 2015. 
	}
\thanks{Asif Mushtaq is with the Department of Mathematical Sciences,
	Norwegian University of Science and Technology (NTNU), Trondheim. e-mail: asif.mushtaq@math.ntnu.no}
\thanks{Trond Kvamsdal is with the Department of Mathematical Sciences,
	NTNU, Trondheim, N-7491, Norway. e-mail: trond.kvamsdal@math.ntnu.no}
\thanks{K{\aa}re Olaussen is with the Department of Physics,
	NTNU, Trondheim, N-7491, Norway. e-mail: Kare.Olaussen@ntnu.no}
}

\maketitle

\pagestyle{empty}
\thispagestyle{empty}

\begin{abstract}

We announce some Python classes
for numerical solution of partial differential
equations, or boundary value problems of
ordinary differential equations.
These classes are built on routines
in \texttt{numpy} and \texttt{scipy.sparse.linalg}
(or  \texttt{scipy.linalg} for smaller problems).

\end{abstract}

\begin{IAENGkeywords}
Boundary value problems,
partial differential equations,
sparse scipy routines.
\end{IAENGkeywords}

%
\IAENGpeerreviewmaketitle

\section{Introduction}

\lstset{
	emphstyle=\color{MyDarkBlue}\bfseries,
	emph={shape,dim,size,bC,geometry,r0,rE,dr,set_bC,
		domain,targetNsource,narr,rvec,qarr,kvec,def_f,def_F,def_g,def_G,
		evalf,evalF,evalg,evalG,evalfn,evalFr,evalgq,evalGk,
		evalFr,values,fftvalues,FFT,iFFT,shift,restrict,prolong,
		lattice,matrix,linOp,varOp,stensOp,laplace,stensil}
}

%
%
%
%

\IAENGPARstart{T}{he} Python computer language has
gained increasing popularity in
recent years. For good reasons: It is fast and easy
to code and use for small ``prototyping'' tasks, since
there is no need for explicit declaration of variables
or a separate compilation cycle. It is freely available
for most computer platforms, and comes with a huge
repository of packages covering a large area of applications.
Python also have features which
facilitates development and encourages documentation
of large well-structured program systems.


Obviously, as an interpreted language native Python is not
suitable for performing extended numerical computations.
But very often the code for such computations reduces
to calls to precompiled library routines.
The \texttt{\bfseries numpy} \cite{Walt_etal} and
\texttt{\bfseries scipy} \cite{Jones_etal, Oliphant} packages
make a large number of such routines directly available
from Python. These packages are freely available for
most operating systems, including {Linux},
{OSX}, and {MSWindows}.

We here describe a process of making some of these 
routines even simpler to use for a field
of applications, the numerical solution of partial
differential equations discretized on a rectangular
grid (or a subdomain of such a grid). As a simple
reference problem one may consider the solution of
the wave equation in the frequency domain,
\begin{equation}
    \left(-\Delta + \omega^2\right)\varphi(\bm{x}) = f(\bm{x}),
\end{equation}
f.i.~in a space with periodic boundary conditions. Our work
is to a considerable extent motivated by a goal
to solve the 3D acoustic wave equation with position
dependent material properties,
and its related inverse problem
\cite{Operto_etal, PederEliasson},
to interesting accuracy in acceptable time on
current (2015) high-end laptops.

However, the classes used to solve this problem are designed
with additional topologies, geometries, and applications in mind.
These classes are \lstinline!Lattice!, \lstinline!LatticeFunction!,
and \lstinline!LatticeOperator!.
A specific application from Quantum Mechanics \cite{AmnaKare} has
been refactored to extend these classes. 


\section{The Lattice class}

This class is intended to handle the most basic properties
and operations of a discretized model. We divide them
into topological and geometrical aspects of the model.
The most basic properties of a dicrete model are the dimensionality
of space, and how we  approximate a continuous space with a number of sites
in each direction (referred to as its \lstinline!shape!). 
The code snippet

\vspace{-4ex}
\begin{lstlisting}
L1 = Lattice(shape=(2**13, ))
L2 = Lattice(bC=('P', 'A'))
L3 = Lattice(shape=(2**8, 2**8, 2**7))
\end{lstlisting}

\noindent
demonstrate how three \lstinline!Lattice! instances
can be defined, \lstinline!L1!
with a one-dimensional lattice of $2^{13} = 8\,192$ sites,
\lstinline !L2! with (by default) a two-dimensional lattice
of $2^7 \times 2^7$ sites,
and \lstinline!L3! with a three-dimensional lattice 
of $2^{8} \times 2^{8} \times 2^{7} = 8\,388\,608$ sites.
In this process the instance properties \lstinline!shape!, \lstinline!dim!,
and \lstinline!size! are specified or given default values.

\subsection{Boundary conditions}

One additional property,  \lstinline!bC!, specifies the default boundary
conditions in all directions. These conditions specify how functions
defined on a finite lattice is extended beyond its
edges, as is required when applying discrete differential
operators or operations like Fast Fourier Transforms (FFT).

Each specific case of \lstinline!bC! is a property
of each function defined on the lattice. Hence it belongs to the class
\lstinline!LatticeFunction!, to be used and set by methods of
\lstinline!LatticeOperator!. However, since \lstinline!bC!
is  often the same for all functions and operators in a
given lattice model, it is convenient to provide a default
property, which may be inherited by instances of
\lstinline!LatticeFunction! and \lstinline!LatticeOperator!.

The default value of \lstinline!bC! is
\lstinline!'allP'!, for periodic boundary conditions
in all directions. 
Otherwise, \lstinline!bC! must be
a list with possible entries
\lstinline!'P'! (for periodic extension),
\lstinline!'S'! (for symmetric extension),
\lstinline!'A'! (for antisymmetric extension),
\lstinline!'F'! (for extension with fixed provided values), 
and \lstinline!'Z'! (for extension with zero values).

For \lstinline!'S'! and \lstinline!'A'! the symmetry point 
is midway between two lattice points.
The boundary condition can be specified
differently in different directions, and (unless
periodic \lstinline!'P'!) differently at the
two edges of a given direction (in which case the
corresponding entry in \lstinline!bC! must be
a two-component list). Internally
\lstinline!bC! is either stored as \lstinline![['allP', ]]!, or
as a \lstinline!dim!-component list of two-component
lists.
The \lstinline!Lattice! class is equipped with a
method, \lstinline!set_bC(bC='allP')!,
which returns the internal re\-pre\-sen\-ta\-tion from a variety
of possible inputs. 

\subsection{Subdomains and slices}

Assume that $\phi(\bm{n})$ and $\phi_{\text{O}}(\bm{n})$ are
two arrays defined on a 3-dimensional lattice, with $s_{\text{O}}$
a constant, and that we want to perform the operation
\begin{equation}
	\phi_{\text{O}}(\bm{n}) = \phi_{\text{O}}(\bm{n})
		+ s_{\text{O}}\,\phi(\bm{n}).
\end{equation}
Python code for this operation could be the snippet

\vspace{-4ex}

\begin{lstlisting}
for nx in range(phi.shape[0]):
	for ny in range(phi.shape[1]):
		for nz in range(phi.shape[2]):
			phiO[nx, ny, nz] = \
				phiO[nx, ny, xz] + \
				sO*phi[nx, ny, nz] 
\end{lstlisting}
This code is lengthy (hence error-prone) and runs
slowly, because all \lstinline!for!-loops are
executed in native Python. The \texttt{\bfseries numpy} code for
the same operation is simply

\vspace{-4ex}

\begin{lstlisting}
phiO += sO*phi
\end{lstlisting}
wherein all loop operations are delegated to \texttt{\bfseries numpy} 
(and maybe further
translated to optimized \texttt{\bfseries BLAS} operations).\footnote{Note
that the codeline 
\lstinline!phiO = phiO + s0*phi!
is \emph{not} equivalent to
\lstinline!phiO += s0*phi!. In the former a
new copy of \lstinline!phiO! is made; this requires
more memory.
}
The similar operation corresponding to
\begin{equation}
    \phi_{\text{O}}(\bm{n}) = \phi_{\text{O}}(\bm{n}) + \sum_{\bm{b}} s_{\text{O}}(\bm{b})\,\phi(\bm{n}-\bm{b}),
\end{equation}
where $\bm{b}$ is a non-zero integer vector, requires
more care and coding, since there will be values
of $\bm{n}$ for which $\bm{n} - \bm{b}$ falls
outside the lattice. In such cases the
expression $\phi(\bm{n}-\bm{b})$ must be
related to known values of $\phi$ by use
of the boundary conditions. Assume a case
where \lstinline!b = (3,0,-2)!,
that the lattice have (much) more
than 3 sites in all directions,
and that the boundary conditions is given by

\vspace{-4ex}

\begin{lstlisting}
bC = [['P','P'],['S','A'],['A','S']]
\end{lstlisting}
We may first treat the sites $\bm{n}$ where also
$\bm{n}-\bm{b}$ fall inside the lattice:

\vspace{-4ex}

\begin{lstlisting}
phiO[3:,:,-2] += sO[3,0,-2]*phi[:-3,:,2:]
\end{lstlisting}
Here the \emph{slice}-notation defines
a rectangular subdomain of the lattice.
For instance, the slice \lstinline![3:,:,-2]! specifies
the intersection of (i) all planes in
the $x$-direction except the first 3,
(ii) all planes in the $y$-direction,
and (iii) all planes in the $z$-direction
except the last 2.

Note that array positions
are counted from zero, with negative numbers
referring to distances from the end.
For a large lattice the above operation would
cover most of the cases, and everything if the
boundary conditions were \lstinline!'Z'! in
all directions.

In our example there are three more regions to be included:
\begin{subequations}
\begin{align}
	0\le n_x < 3,&\text{ and }  0 \le n_z < -2,\label{regionI}\\
	3 \le n_x \le -1,&\text{ and }  -2 \le n_z \le -1,\label{regionII}\\
	0 \le n_x < 3,&\text{ and }  -2 \le n_z \le -1.\label{regionIII}
\end{align}
\end{subequations}
The case~\eqref{regionI} can be handled by the code

\vspace{-4ex}

\begin{lstlisting}
phiO[:3,:,:-2] += \
		sO[3,0,-2]*phi[-3:,:,2:]
\end{lstlisting}
using the periodic boundary condition in
the $x$-direction. For the cases \eqref{regionII} and \eqref{regionIII}
two planes in the $z$-direction fall outside the lattice on
the upper side.  Due to the
symmetric \lstinline!'S'! boundary condition at
this edge of the lattice, the function values on these planes
are related to their values on the last two planes inside the
lattice (counted in opposite order).
This can be handled by the code


\begin{lstlisting}
phiO[3:,:,-2:] += \
		sO[3,0,-2]*phi[:-3,:,:-3:-1]
phiO[:3,:,-2:] += \
		sO[3,0,-2]*phi[-3:,:,:-3:-1]
\end{lstlisting}
For detailed information about indexing and slicing
in \texttt{\bfseries numpy}, consult the \emph{Indexing} section of
the Numpy reference manual \cite{NumpyReference}.
However, gory details like the above are
best handled by computers. The \lstinline!Lattice! class
provides a method,
\lstinline!targetNsource(b, bC=None)!,
which yields all the source and target slices required
for a given vector $\bm{b}$.
Using this, the code snippet

\vspace{-4ex}

\begin{lstlisting}
for cf, dT, dS in L3.targetNsource(b):
	phiO[dT] += cf*phi[dS]
\end{lstlisting}
replaces all operations above. Here the coefficient \lstinline!cf!
is $-1$ if an odd number of antisymmetric
boundary conditions are employed (otherwise $+1$).

\lstinline!Lattice! also provides a
related method,
\lstinline!domain(shape, shift)!.
This returns a slice \lstinline!dI! pointing
to a rectangular subdomain of the lattice, of
shape \lstinline!shape!, shifted from the origin
by an integer vector \lstinline!shift!.

\subsection{Index arrays and broadcasting}

Each site of a \lstinline!dim!-dimensional lattice is labeled by
a \lstinline!dim!-dimensional integer index vector $\bm{n}$.
To construct an array \lstinline!A! defined on all points of a 3-dimensional lattice, one could write a code snippet similar to the following

\vspace{-4ex}

\begin{lstlisting}
defA = lambda n: \
			numpy.exp(-numpy.dot(n,n))
shape = (2**8, 2**8, 2**8)
A = numpy.zeros(shape)
for n in numpy.ndindex(shape):
	A[n] = defA(numpy.array(n))
\end{lstlisting}
Although this code is brief and general with respect to dimensionality,
it is \emph{not a good way to do it}. Since the
\lstinline!for!-loop will be executed in native Python, the code
will run too slow. A better way is to define three arrays
\lstinline!n0!, \lstinline!n1!, \lstinline!n2!, all
of shape $(2^8, 2^8, 2^8)$, once and for all.
We may then replace the code above with the snippet
\vspace{-4ex}

\begin{lstlisting}
defA = lambda n0, n1, n2: \
	exp(-n0*n0)*exp(-n1*n1)*exp(-n2*n2)
A = defA(n0, n1, n2)
\end{lstlisting}

\noindent
All loops are now implicit, and will be executed by 
compiled \texttt{\bfseries numpy} functions.

Further, the memory cost of permanently storing
three large arrays can be avoided by use of the \emph{broadcasting}
facility of \texttt{\bfseries numpy}.
Since the index array \lstinline!n0! is constant
in the $y$- and $z$-directions, it only contains
a one-dimensional amount of information, stored in
an array of shape $(2^8, 1, 1)$. Likewise, \lstinline!n1!
can be stored in an array of shape $(1, 2^8, 1)$, and \lstinline!n2!
in an array of shape $(1,1,2^8)$. All these arrays contain the same
amount of data ($2^8$ linearely stored entries). But, due to their
different \emph{shape} they will act differently under f.i. algebraic
operations: \lstinline!n0*n0! will still produce an array of shape
$(2^8,1,1)$, and similary \lstinline!n1*n1! an array of shape $(1,2^8,1)$.
However, the addition of these two results produces an array of shape
$(2^8, 2^8, 1)$. 

Finally, adding \lstinline!n2*n2! generates an array
of the final shape $(2^8, 2^8, 2^8)$.
Hence, the cost of computing and storing index arrays are
modest. We have chosen \emph{not} to include them
as properties, but provide a method \lstinline!narr()! which computes
them when needed. This method returns a
{list} \lstinline!n! of arrays, \lstinline![n[0], n[1],..]!.

\subsection{Geometric properties}

The discussion above maily concerns topological properties of the lattice.
For most application we also need some geometric properties. In general these
may be implemented by defining a \lstinline!dim!-dimensional vector of arrays,
$\bm{r(n)}$, specifying the position coordinates of all sites. These
coordinates (which should depend monotoneously on $\bm{n}$)
could also be dynamical, i.e.~part of the equation system to be
solved.

The wide range of possibilities indicate
that several versions of $\bm{r(n)}$
should be implemented, with the appropriate version
chosen when a \lstinline!Lattice! instance is defined.
We have introduced a property \lstinline!geometry!,
which specifies the version to be used.
So far, \lstinline!geometry! can only take the value \lstinline!'fixedRect'!,
wherein rectangular regions of space, aligned with the
lattice directions, are modelled. Such regions can be specified
by a \lstinline!dim!-dimensional vector $\bm{r}_E$ of edge-lengths,
plus a vector $\bm{r}_0$ specifying
the position of the ``lower left'' corner of the spatial region. For a
given lattice \lstinline!shape! parameter, this defines
a lattice cell with a vector of sidelengths $\bm{dr}$, such that
\begin{equation}
   \text{\lstinline!dr[d] = rE[d]/shape[d]!}.
   \label{dr}
\end{equation}
The position coordinate $\bm{r(n)}$ is then defined
such that its component in the \lstinline!d!-direction is
\begin{equation}
   \text{\lstinline!r[d] = r0[d] + dr[d]*(n[d]+1/2)!}.
   \label{r_n}
\end{equation}
This implementation introduces three new properties: 
\lstinline!r0!, by default a \lstinline!dim!-dimensional
tuple with entries $0$,
\lstinline!rE!, by default a \lstinline!dim!-dimensional
tuple with entries $1$,
and \lstinline!dr!, calculated from equation~\eqref{dr}.
The method \lstinline!rvec()! returns a list \lstinline!r!
of arrays, \lstinline![r[0], r[1],...]!,
calculated from equation~\eqref{r_n}.

\subsection{Lattice initialization, methods, and properties}

All currently available keyword arguments and default values
for initialization of a \lstinline!Lattice! instance is specified
by the code snippet below:

\vspace{-4ex}

\begin{lstlisting}    
def __init__(shape=(128, 128),
	bC='allP', geometry='fixedRect',
	rE=None, r0=None):
\end{lstlisting}
A value of \lstinline!None! will invoke a default initialization
process, following the rules discussed above.
The current list of \lstinline!Lattice! methods, with
arguments, is as follows:

\begin{center}
\begin{tabular}{rl}
\lstinline!set_bC!&\lstinline!(bC='allP')!\\
\lstinline!domain!&\texttt{(shape, shift)}\\
\lstinline!targetNsource!&\lstinline!(b, bC=None)!\\
\lstinline!narr!&\lstinline!()!\\
\lstinline!rvec!&\lstinline!()!
\end{tabular}
\end{center}

A summary of all \lstinline!Lattice! properties, with example values, is
as provided in the table below.


\begin{center}

\begin{tabular}{rl}
\lstinline!shape!& \lstinline!L1.shape = (8192, )!\\
&\lstinline!L3.shape = (256, 256, 128)!\\
\lstinline!dim!& \lstinline!L1.dim = 1!\\
&\lstinline!L3.dim = 3!\\
\lstinline!size! & \lstinline!L1.size = 8192!\\ 
&\lstinline!L3.size = 8388608!\\
\lstinline!bC! & \lstinline!L1.bC = [['allP']]!\\
&\lstinline!L2.bC = [['P','P'],['A','A']]!\\
\lstinline!geometry! & \lstinline!L1.geometry = 'fixedRect'!\\
\lstinline!r0! & \lstinline!L1.r0 = (0, )!\\
&\lstinline!L3.r0 = (0, 0, 0)!\\
\lstinline!rE! &\lstinline!L1.rE = (1, )!\\
&\lstinline!L3.rE = (1, 1, 1)!\\
\lstinline!dr! &\lstinline!L1.dr = [1.19209290e-07]!\\
&\lstinline!L3.dr = [0.00390625,!\\
&\qquad\qquad\lstinline!0.00390625, 0.0078125]!
\end{tabular}

\end{center}

\section{The LatticeFunction class}

Space does not allow us to continue with an equally
detailed discussion of all components in the
\lstinline!LatticeFunction! and \lstinline!LatticeOperator!
classes. We will instead provide examples of uses,
augmented with general comments.

\vspace{-4ex}

\begin{lstlisting}
    L = Lattice(shape=(2**15, 2**15),
    rE=(18, 18), r0=(-9, -9))
    defF = lambda r: \
		numpy.exp(-r[0]**2/2)* \
		numpy.exp(-r[1]**2/2)
    F = LatticeFunction(L, def_F=defF)
    t0 = time.time()
    F.evalFr()
    print (L.size,
    (time.time()-t0)/L.size)
\end{lstlisting}
Here we first define a $2^{15} \times 2^{15}$ lattice model
with periodic boundary conditions, and next a gaussian function
centered in the middle of this lattice. The parameter \lstinline!rE!
is chosen large enough to make the periodic extension of this
function smooth: It acquires a discontinuity in the first
derivative of magnitude $18 \exp(-9^2/2) \approx 0.5\cdot 10^{-16}$
or smaller (i.e., below double precision accuracy).

The gaussian function is
\emph{not} evaluated when the instance \lstinline!F!
is defined, only when we execute the method \lstinline!F.evalFr()!.
This method evaluates the function, and stores the result in
the array \lstinline!F.values!.

The wall-clock time used to perform this
computation on a 2013 MacBook Pro with 16 Gb of memory
was measured to $3.26\;\text{ns}$ per point. Note that this time is
mostly spent multiplying double precision numbers;
only $2\times 2^{15}$ exponential function evaluations
are performed. However,
if the the code for \lstinline!def_F! is changed to

\begin{figure}[h]
\includegraphics{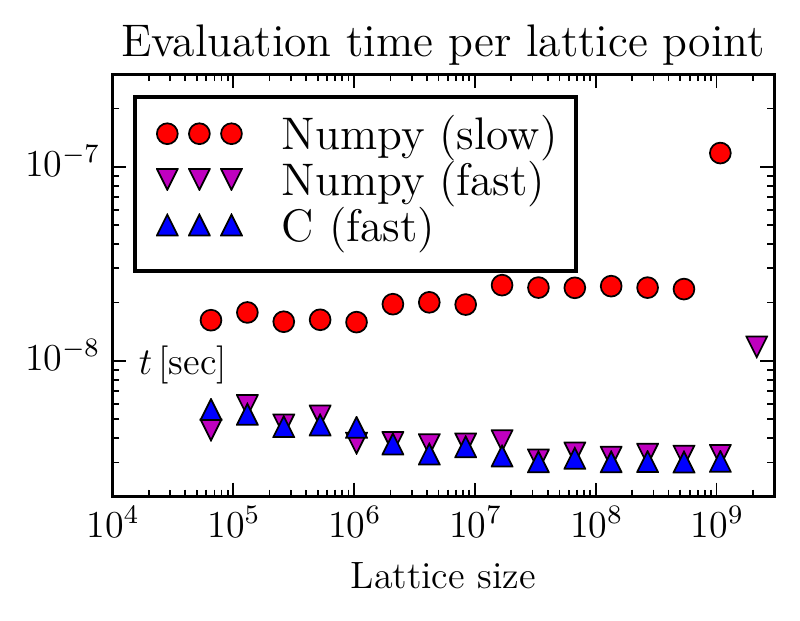}
\vspace{-3ex}
\caption{\label{figure0}
Comparison of NumPy and C evaluation times for a gaussian defined on
lattices of various sizes. The fast evaluation occur when writing the gaussian
as $\exp(-x^2/2) \times \exp(-y^2/2)$, the slow evaluation when writing
it as $\exp[-(x^2 + y^2)/2]$. For these cases the wall-clock and CPU times
are essentially the same.
As can be seen, there is little to gain in evaluation time by writing
the code in a fast, compiled language like C (and a lot to lose in
coding time).
}
\end{figure}

\vspace{-4ex}

\begin{lstlisting}
def_F = lambda r: \
	numpy.exp(-(r[0]**2 + r[1]**2)/2)
\end{lstlisting}
the execution time increases to $118\;\text{ns}$ per point.
This increase is partly due to the fact that the exponential
function is now evaluated $10^{30}$ times, but also
because the system now has to deal with \emph{two}
very large arrays (one for the argument of
the exponential function, and one for final result),
and is operating very close to the limit of available
memory. A more detailed analysis, for lattices of various
sizes (total number of lattice points), is shown in Fig~\ref{figure0}.

\vspace{-4ex}

\subsection{FFT and related discrete transforms}

We may apply a discrete Fourier transformation to the
data stored in \lstinline!F.values!. This is done by
the function call \lstinline!F.FFT()!. The
transformed data is stored in the
array \lstinline!F.fftvalues!. The inverse transform
is performed by the function call \lstinline!F.iFFT()!,
with the transformed data being stored in the
array \lstinline!F.values! (overwriting any previous data).

Acctually, the method \lstinline!FFT()! (or
\lstinline!iFFT()!) do not necessarily perform
a regular (multidimensional) discrete Fourier transform 
\lstinline!fftn! (or its
inverse \lstinline!ifftn!).
This is but one of several related discrete transforms available in
\texttt{\bfseries scipy.fftpack}. Other such transforms
are the discrete cosine transform \lstinline!dct! (suitable
for functions with symmetric boundary conditions on
both sides), the discrete sine transform \lstinline!dst! (suitable
for functions with antisymmetric boundary conditions on both
sides), the fast fourier transform \lstinline!rfft! of real data, and
their inverses (\lstinline!idct!, \lstinline!idst!, \lstinline!irfft!).
The rules are
\begin{enumerate}

\item
If \lstinline!bC[0][0] == 'allP'! the transform
\lstinline!fftn!
(or \lstinline!ifftn!) is used. Complex data is allowed.
Otherwise, the data is assumed to be real, and an
iterated sequence of transforms over all axes is
executed.

\item
For directions such that \lstinline!bC[d][0] == 'S'! the transform
\lstinline!dct! (or \lstinline!idct!) is performed.

\item
For directions such that \lstinline!bC[d][0] == 'A'! the transform
\lstinline!dst! (or \lstinline!idst!) is performed.

\item
In all other cases the transform \lstinline!rfft! 
(or \lstinline!irfft!) is performed.

\end{enumerate}

Note that the \lstinline!bC! used here is a property of \lstinline!LatticeFunction!.
This may be different from the corresponding property of
its lattice instance. By default they are equal.

The discrete transforms above are useful because they allow (i)
differential operators to be implemented as
multiplication operators on the transformed functions, and
(ii) accurate interpolation of lattice functions outside
the lattice sites. The latter is useful for implementation of
prolongations in multigrid methods. To assess to which extent this
is a practical approach, we have investigated the accuracy and the time
requirements of these transforms. The code snippet below illustrate
how this can be done:

\vspace{-4ex}

\begin{lstlisting}
shape = (2**14, 2**14)
L = Lattice(shape=shape, bC=('P','P'))
F = LatticeFunction(L)
F.values = numpy.random.rand(*shape)
values = numpy.copy(myF.values)
t0 = time.time(); F.FFT()
t1 = time.time(); F.iFFT()
t2 = time.time()
err=numpy.max(numpy.abs(F.values-values))
print((t1-t0)/L.size,(t2-t1)/L.size,err)
\end{lstlisting}

The output of this code shows that the forward transform
takes about $68\,\text{ns}$ per lattice point, the
inverse transform about $57\,\text{ns}$, and that the
maximum difference between the original and backtransformed
values is $1.7\times 10^{-15}$. I.e., the cost of a one-way
transform is roughly the same as 20 multiplications.
The time per site increases
by almost an order of magnitude for
a lattice of \lstinline!shape = (2**14, 2**15)!, since this is close
to the limit of available memory. 

\begin{figure}[h]
\includegraphics{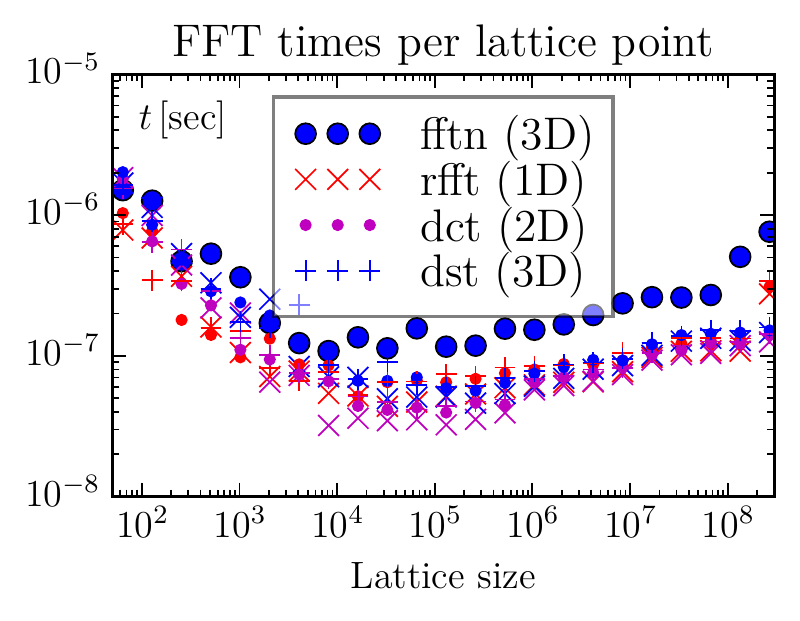}
\vspace{-4ex}
\caption{\label{figure1}
Time used to perform a discrete lattice transformations
of various types. Each time plotted is the \emph{sum} of the
forward and inverse transformation time. 1D lattices are
plotted in red, 2D lattices in magenta, and 3D lattices
in blue. Within the range of lattice sizes allowed by available memory,
the theoretically expected logarithmic growth of transformation time 
with size is not a very distinct feature.
}
\end{figure}

We have investigated the behavior above in more detail, for
different choices of the \lstinline!shape! and \lstinline!bC!
parameters, with similar results. See Fig.~\ref{figure1}.
The crude conclusion is that the transformation times
grow roughly linearly with lattice size, with a prefactor
which depends only slightly on transformation type and
lattice dimensionality.

\vspace{-4ex}

\subsection{LatticeFunction initialization, methods, and properties}

All currently available keyword arguments for initialization of a
\lstinline!LatticeFunction! instance is specified by the argument list below:

\vspace{-4ex}

\begin{lstlisting}
def __init__(self, lattice, def_f=None,
	def_F=None, def_g=None, def_G=None,
	bC='allP', evalf=False, evalF=False,
	evalg=False, evalG=False):
\end{lstlisting}
As can be inferred from the above, there are several ways to
specify a function: (i) As a function of the index arrays,
$f(\bm{n})$, or as a function of the position vectors,
$F(\bm{r})$. The discrete transform of the function also lives
on a lattice, the \emph{dual lattice}, whose sites
can be labelled by a list of
index arrays \lstinline!q = [q[0], q[1],..]!. We denote
the geometric version of this lattice as
\emph{reciprocal space}, wherein each site $\bm{q}$ has
a reciprocal position vector $\bm{k}(\bm{q})$. Hence,
the function can also be specified from its discrete
transformation, as the function (iii) $g(\bm{q})$ or
(iv) $G(\bm{k})$.

The current list of \lstinline!LatticeFunction! methods is as follows:

\begin{center}
\begin{tabular}{rl}
\lstinline!qarr()!   & List index vectors for the dual lattice.\\
\lstinline!kvec()!   & List resiprocal position vectors.\\
\lstinline!evalfn()! & Compute \lstinline!values! from \lstinline!def_f!.\\
\lstinline!evalFr()! & Compute \lstinline!values! from \lstinline!def_F!.\\
\lstinline!evalgq()! & Compute \lstinline!fftvalues! from \lstinline!def_g!.\\
\lstinline!evalGk()! & Compute \lstinline!fftvalues! from \lstinline!def_G!.\\
\lstinline!FFT()!    & Discrete transformation of \lstinline!values!.\\
\lstinline!iFFT()!   & Inverse transformation of \lstinline!fftvalues!.\\
\lstinline!shift(frac)! & Return the
function translated by \lstinline!frac!.\\
\lstinline!restrict()! & Return the function restricted\\
&	to a cruder lattice.\\
\lstinline!prolong()! & Return the function prolonged\\
& to a finer lattice.
\end{tabular}
\end{center}

The current list of \lstinline!LatticeFunction! properties is as follows:

\begin{center}
\begin{tabular}{rl}
\lstinline!lattice!  & Related \lstinline!Lattice! instance.\\
\lstinline!def_f!    & Possible function definition (default \lstinline!None!).\\
\lstinline!def_F!    & Possible function definition (default \lstinline!None!).\\
\lstinline!def_g!    & Possible function definition (default \lstinline!None!).\\
\lstinline!def_G!    & Possible function definition (default \lstinline!None!).\\
\lstinline!bC!       & Boundary conditions (\lstinline!lattice.bC!).\\
\lstinline!values!   & Array of function values.\\
\lstinline!fftvalues!& Array of transformed function values.
\end{tabular}
\end{center}

\section{The LatticeOperator class}

Many routines in \texttt{\bfseries scipy.sparse.linalg}
do not require an explicit matrix representation of
the operator under analysis. Only some algorithm which returns
the result of applying the operator to a given vector
is needed. Such algorithms can be assigned to a
\lstinline!LinearOperator! instance, after which it functions
essentially as an explicit matrix representation.
Such algorithms should not demand too much memory
or computation time, but do not require any explicitly known
sparse representation of the operator. F.i., any computational
process involving a fixed number of multiplication, additions
and fast fourier transformations will have a memory requirement
which scales linearly with the lattice size, and a time
requirement which (for large systems) also scales roughly linearly with lattice
size.

The \lstinline!LinearOperator! class requires an input vector of shape
\lstinline!(M, )! or \lstinline!(M,1)!, and an output vector
of shape \lstinline!(N, )!. For higher-dimensional lattices
this does not match the natural construction of
lattice operators, which we do not want to interfere
with. We have therefore implemented a general
\lstinline!linOp(phi0)! method, to be used as a
universal \lstinline!matvec! parameter for \lstinline!LinearOperator!.
The currently implemented code for this is

\vspace{-4ex}

\begin{lstlisting}
phi = phi0.reshape(self.lattice.shape)
return numpy.ravel(self.varOp(phi)) 
\end{lstlisting}
This code assumes \lstinline!phi0! to represent a
scalar function. It will be extended to
more general (vector, spinor, tensor,\ldots)
objects. The \lstinline!reshape! and \lstinline!ravel!
operations above do not modify
or move any data; they only change how
the data is interpreted (the \emph{view} of the data).

The code above also call a specific method, 
\lstinline!varOp(phi)!. However,
this is just a handle which should be assigned
to the operator under analysis. The latter may
either be an appropriate predefined method
in the \lstinline!LatticeOperator! class, or
a method provided from outside.

\subsection{Explicit matrix representations}

It may be useful to inspect an explicit matrix
representation of a given operator on a small
lattice. The method \lstinline!matrix(operator)!
provides such a representation:

\vspace{-4ex}
 
\begin{lstlisting} 
L = Lattice(shape=(4, ), rE=(4,))
O = LatticeOperator(L)
laplace = O.matrix(O.laplace)
print (laplace)
\end{lstlisting}
The output from this code is

\vspace{-4ex}

\begin{lstlisting}
[[-2.  1.  0.  1.]
[ 1. -2.  1.  0.]
[ 0.  1. -2.  1.]
[ 1.  0.  1. -2.]]
\end{lstlisting}
which is easily verified to have the
correct form for a $3$-stensil one-dimensional
lattice Laplacian with periodic
boundary conditions. We may redefine
the lattice to have the \lstinline!'Z'!
boundary condition:

\vspace{-4ex}

\begin{lstlisting}
L = Lattice(shape=(4,), bC='Z', rE=(4,))
\end{lstlisting}
The output now becomes:

\vspace{-4ex}

\begin{lstlisting}
[[-2.  1.  0.  0.]
[ 1. -2.  1.  0.]
[ 0.  1. -2.  1.]
[ 0.  0.  1. -2.]]
\end{lstlisting}
When applied to a small two-dimensional lattice

\vspace{-4ex}

\begin{lstlisting}
L = Lattice(shape=(2,3),bC='Z',rE=(2,3))
\end{lstlisting}
the output for the correponding $5$-stensil becomes

\vspace{-4ex}

\begin{lstlisting}
[[-4.  2.  0.  0.  0.  0.]
[ 2. -4.  2.  0.  0.  0.]
[ 0.  2. -4.  0.  0.  0.]
[ 0.  0.  0. -4.  2.  0.]
[ 0.  0.  0.  2. -4.  2.]
[ 0.  0.  0.  0.  2. -4.]]
\end{lstlisting}
We have found such applications of the \lstinline!matrix()!
method to be quite educating, and very useful for debugging
purposes.

The output matrix can also be used directly
as input to all the standard (dense matrix) linear algebra
routines in \texttt{\bfseries scipy}. Lattice sizes up to about $10^4$
can be handled in this way, sufficient for
most one-dimensional systems (and useful when comparing
dense and iterative methods on small higher-dimensional
systems).

Methods for generating sparse matrix
representations will also be implemented.

\vspace{-2ex}

\subsection{Example of use}

\begin{figure}[b]
\includegraphics{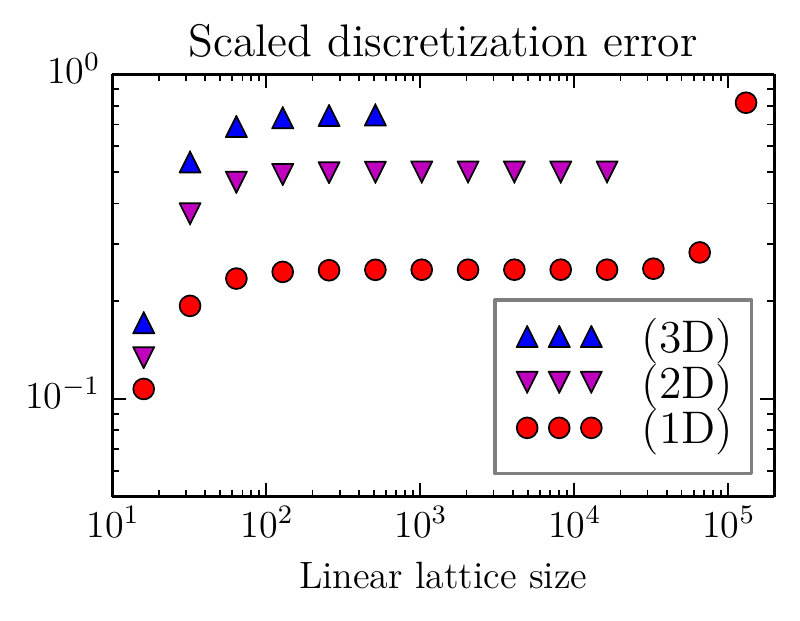}
\vspace{-3ex}
\caption{\label{deltaLaplace}
The maximum absolute difference between the
numerical and exact evaluation of the Laplace operator,
divided by $dr^2$,
as function of the linear lattice size. This shows that
the error scales like $dr^2$, as expected for these
stensils. The increase in error for large linear size
is probably due to numerical roundoff (because $dr^2$ becomes
very small), the decrease for
small linears size due to incomplete sampling of errors (too few
lattice points to compare the functions
where the error is maximum).
}
\end{figure}

An example illustrating the discussion above is provided
by the code snippet:

\vspace{-4ex}

\begin{lstlisting}
L = Lattice(shape=(2**8, ),
		rE=(18, ), r0=(-9, ))
defF = lambda r: numpy.exp(-r[0]**2/2)
F = LatticeFunction(L, def_F=defF,
		evalF=True)
O = LatticeOperator(L)
O.varOp = O.laplace
F2values = O.varOp(F.values)
\end{lstlisting}

In this simple case it does not matter if \lstinline!F2values!
is computed by use of \lstinline!O.linOp!,
\lstinline!O.varOp! or \lstinline!O.laplace!. The result of
evaluating $\Delta_L \exp(-r^2/2)$ can be compared with the exact result,
$(r^2-1)\,\exp(-r^2/2)$. 
A good way to assess the discretization error is to
compute $\max_{\bm{r}} \left| \Delta_L F(\bm{r}) - 
\Delta F(\bm{r})\right|$. This is plotted in Fig.~\ref{deltaLaplace}
for a range of square lattices.


\vspace{-2ex}

\subsection{The lattice Laplace operator}

\begin{figure}[h]
\includegraphics{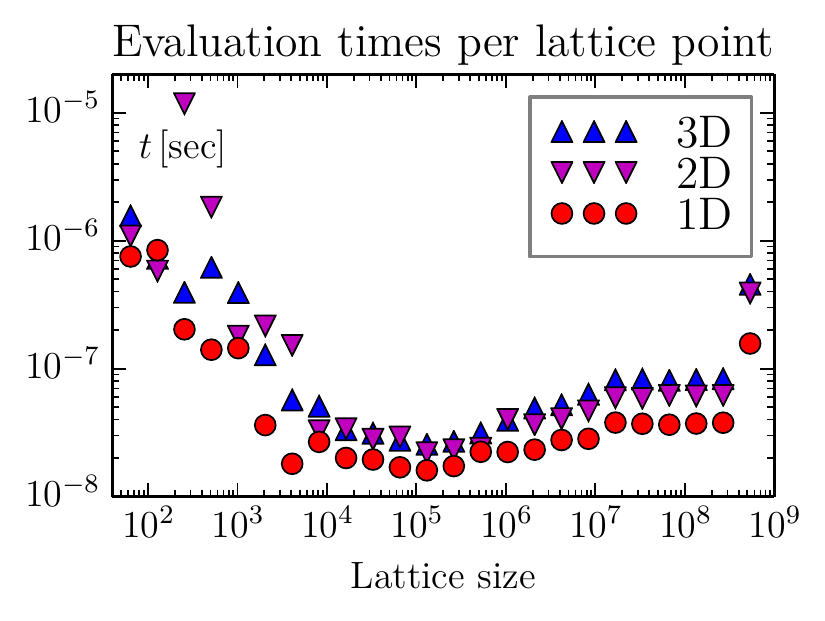}
\vspace{-3ex}
\caption{\label{figure2}
The times to evaluate $-\Delta_L \phi$,
for the (standard) $(2D+1)$-sensil approximation
of the Laplace operator, are plotted for various
lattice sizes and dimensionalites. As expected,
the times increases with the complexity of the
stensil. Somewhat surprisingly, the times are not
significantly different from the times to
perform back-and-forth fast fourier transform
(or its discrete analogs), c.f.~Fig.~\ref{figure1}. 
}
\end{figure}

We have used a simple implementation of the lattice Laplacian in
the examples above.
This is the common $(2d +1)$-stensil approximation.
For periodic boundary conditions the implemention is very simple,
as indicated by the code snippet below:

\vspace{-4ex}

\begin{lstlisting}
def laplace(self, phi):
	Lphi = numpy.zeros_like(phi)
	for d in range(self.dim):
		Lphi += numpy.roll(phi, 1, 
			axis=d)
		Lphi += numpy.roll(phi,-1, 
			axis=d)
		Lphi -= 2*phi
	return Lphi/self.dr**2
\end{lstlisting} 

Here the \lstinline!roll!-function rotates the entries
of the \lstinline!phi!-array in the \lstinline!d!-direction by the
specified amount ($\pm 1$ for the code above).
We have investigated how fast this implementation is.. The
results is plotted in Fig.~\ref{figure2}. As expected, the
evaluation times scales (essentially) linearly will lattice
size, with a prefactor which increases with the complexity
of the stensil. But, somewhat surprisingly, the evaluation
times are not very different from the time to make back-and-forth
fast fourier fourier transformations.
This suggests an alternative approach,
based on fast fourier transforms.

The roll-process is fast, with all loop operations
done in NumPy, but requires new memory for the rolled data.
To avoid this we have implemented a general method,
\lstinline!stensOp(phi)!. The essential
algorithm of this is illustrated by the snippet below:

\vspace{-4ex}

\begin{lstlisting}
for b in numpy.ndindex(stensil.shape):
	cf, dT, dS = lattice.targetNsource(b)
	phiO[dT] += cf*stensil[b]*phi[dS]
\end{lstlisting}
Here \lstinline!stensil! is a (small)
\lstinline!dim!-dimensional array
defining the operator in question.

\vspace{-2ex}

\section*{Acknowledgment}

We thank dr.~Peder Eliasson (Research manager, SINTEF petroleum Trondheim) for an informative discussion.
This work has been partially supported by the UniCQue project.

\end{document}